\documentclass[twocolumn,trackchanges]{aastex631}

\usepackage{newtxtext,newtxmath}

\usepackage{graphicx}	
\usepackage{amsmath}	
\usepackage{amssymb}	
\usepackage{gensymb}
\usepackage{xcolor}
\usepackage{multirow}

\usepackage{tikz,xcolor,hyperref}

\shorttitle{Diffuse Emission in A1351}
\shortauthors{S. Chatterjee et al.}

\graphicspath{{./}{figures/}}

\defcitealias{Giacintucci_2009ApJ...704L..54G}{G09}
\defcitealias{Barrena_2014MNRAS.442.2216B}{B14}

\begin{document}

\title{Unveiling The Origin of Peculiar Diffuse Radio Emission in Abell 1351}

\correspondingauthor{Swarna Chatterjee}
\email{chatterjee.swarna1605@gmail.com}

\author[0000-0001-8194-8714]{Swarna Chatterjee}
\affiliation{Department of Astronomy Astrophysics and Space Engineering, Indian Institute of Technology Indore, Indore, M.P., India}

\author[0000-0002-1372-6017]{Majidul Rahaman}
\affiliation{Department of Astronomy Astrophysics and Space Engineering, Indian Institute of Technology Indore, Indore, M.P., India}
\affiliation{Institute of Astronomy, National Tsing Hua University, Hsinchu, Taiwan}

\author[0000-0002-5333-1095]{Abhirup Datta}
\affiliation{Department of Astronomy Astrophysics and Space Engineering, Indian Institute of Technology Indore, Indore, M.P., India}

\author[0000-0001-8721-9897]{Ramij Raja}

\affiliation{Department of Astronomy Astrophysics and Space Engineering, Indian Institute of Technology Indore, Indore, M.P., India}

\begin{abstract}

Abell 1351 is a massive merging cluster that hosts a giant radio halo and a bright radio edge blended in the halo. In this paper, we present the first ever spectral analysis of this cluster using GMRT 610 MHz and VLA 1.4 GHz archival data and discuss the radio edge property. Using \textit{Chandra} data, we report the first tentative detection of shock front at the location of the edge in A1351 with discontinuities in both X-ray surface brightness and temperature. Our analysis strengthens the previous claim of the detected "edge" being a high luminosity radio relic. The radio relic has an integrated spectral index $\alpha = -1.63 \pm 0.33$ and radio power $P_{1.4 \mathrm{\ GHz}} = 4.46 \pm 0.61\times10^{24}$ W $Hz^{-1}$ with a largest Linear Size (LLS) of 570 kpc. The radio spectral index map shows steepening in the shock downstream region. Our analysis favours the scenario where the Diffusive Shock Acceleration (DSA) of particles is responsible for the origin of the radio relic in the presence of strong magnetic field. We have also estimated the magnetic field at the relic location assuming equipartition condition. 

\end{abstract}


\section{Introduction} \label{sec:intro}
Galaxy clusters play a significant role in the study of the large-scale particle acceleration process. These clusters are a concoction of dark matter, galaxies, and hot gas. The hot gas in the Intra-Cluster Medium (ICM) takes upto 15 - 17\% of cluster total mass (\citealt{vanweeren_2019SSRv..215...16V}), and it emits in the X-ray band through the thermal bremsstrahlung process. Thus X-ray observations provide crucial information about cluster mass and dynamical state \citep{Sarazin_2002}. The gravitational potential energy released during cluster mergers heats the ICM and is channeled through shocks and turbulence in the ICM. These shocks and turbulence accelerate relativistic particles and amplify magnetic fields. As a result, we get large-scale non-thermal synchrotron emissions from clusters in radio wavelength. These diffuse radio emissions come from highly relativistic particles spiraling around cluster magnetic fields and generally show a steep integrated spectral index ($\alpha<-1$ , where $S_{\nu} \propto \nu^{-\alpha}$) \citep{vanweeren_2019SSRv..215...16V}. The emissions are categorised as halo, minihalo, relic, and phoenix \citep{Feretti_2012A&ARv..20...54F,vanweeren_2019SSRv..215...16V}. Halos and minihalos are centrally located diffuse structures. Halos are formed in the Megaparsec scale (size $\sim$ 0.5 - 2 Mpc) in massive merging clusters as a result of turbulent re-acceleration of relativistic electrons in the ICM \citep{Brunetti_2014IJMPD..2330007B}. In comparison, minihalos (size $\sim$ 100 - 500 kpc) are seen in relaxed and cool core clusters mainly and are formed due to minor merger or gas sloshing or due to AGN feedback \citep{ZuHone_2013, Raja_2020ApJ...889..128R, Richard_Laferri_re_2020}. The cluster radio relics are found mainly at the cluster outskirts or peripheral region \citep{Enblin_refId0}, generally have an elongated shape, and are polarised at high frequency. The relics can have the largest linear size (LLS) extending upto 3 Mpc \citep{Hoang_2021arXiv210600679H}. The relics often coincide with cluster merger shocks. This spatial coincidence backs up the idea that relics are shock-generated. The formation of relics, for some clusters, is supported by the theory of Diffusive Shock Acceleration (DSA) of particles where the particles are accelerated to relativistic speed at the shock front from the thermal pool of Intra-Cluster Medium (ICM) \citep{Drury_1984AdSpR...4b.185D, Ensslin_1998A&A...332..395E, Hoeft_2007MNRAS.375...77H, Kang_2013ApJ...764...95K}. However, the DSA mechanism is questioned by the particle acceleration efficiency of weak shocks that are generally observed in clusters \citep{Hoeft_2007MNRAS.375...77H, Botteon_2016MNRAS.460L..84B, Gennaro_2018}. The reported relic luminosities and Mach numbers inferred from radio and X-ray observations point to another possibility where the shock re-accelerates the slightly relativistic old fossil plasma existing in the ICM. This theory is consistent with weaker shocks for generating luminous radio relics \citep{Botteon_2016MNRAS.460L..84B,vanWeeren_2017NatAs...1E...5V, Stuardi_2019MNRAS.489.3905S,rajpurohit2020A&A...636A..30R}.
In another scenario, the shock adiabatically compresses an old AGN bubble and re-energizes fossil plasma. This phenomena is responsible for the generation of radio phoenixes at smaller cluster-centric distance  (\citealt{Enblin_refId0,Slee_2001AJ....122.1172S,ensslin_bruggen2002_10.1046/j.1365-8711.2002.05261.x,Kempner:2003eh,Ferrari:2008jr}). These radio phoenixes have a rather roundish and filamentary morphology, mostly show ultra-steep spectra with $\alpha < -2$ and also hint of spectral curvature at high-frequency \citep{Slee_2001AJ....122.1172S, Kale_2018MNRAS.480.5352K,Mandal_2019A&A...622A..22M}. Despite the available multi-wavelength study, exact particle acceleration mechanisms in the formation of diffuse radio emission are still to be understood properly. Also, we need to increase observational evidence to establish the correlations between cluster properties and diffuse emission.\\
\textbf{Abell 1351} (A1351) is a massive ($M = 1.4 - 4.2 \times 10^{15}\ \mathrm{M}_{\odot}$) merging cluster at a redshift 0.325 \citep[][hereafter \citetalias{Barrena_2014MNRAS.442.2216B}]{Barrena_2014MNRAS.442.2216B}. It has a mass distribution extended along the north-northeast (N-NE) and south-southwest (S-SW) direction (\citealt{Bohringer_2000,Dahle_2002ApJS..139..313D}, \citetalias{Barrena_2014MNRAS.442.2216B}). The presence of diffuse radio emission in A1351 was first detected by \citealt{Owen1999}. Using Karl G. Jansky Very Large Array (VLA) 1.4 GHz data \citealt{Giacintucci_2009ApJ...704L..54G, giovannini_2009A&A...507.1257G} classified this Mpc scale diffuse emission in the cluster as giant radio halo. The halo is also detected by \citealt{Botteon2022} using LOFAR Two-meter Sky Survey Data Release 2 (LoTSS-DR2). As reported previously, the cluster has an X-ray luminosity of $L_x(0.1-2.4\ \mathrm{keV}) = 8.31\times10^{44}\ \mathrm{h}_{50}^{-2}\  \mathrm{ergs}^{-1}$ \citep{Bohringer_2000} and a radio luminosity $P_{1.4\mathrm{\ GHz}} = 1.2-1.3\times10^{25}$ h$^{-2}_{70}$ W Hz$^{-1}$ \citep{Giacintucci_2009ApJ...704L..54G,giovannini_2009A&A...507.1257G}. The diffuse emission shows a bright edge at radio wavelength, which was classified as a "ridge" by \citealt{Giacintucci_2009ApJ...704L..54G} (hereafter \citetalias{Giacintucci_2009ApJ...704L..54G}). Previously bright radio edge blended with the halo emission has been noticed in a few clusters (eg. \citealt{Brown_2011,Macario_2011,Shimwell_2014MNRAS.440.2901S,Wang_2018}). Both \citetalias{Giacintucci_2009ApJ...704L..54G} and \citetalias{Barrena_2014MNRAS.442.2216B} suggested the complex, diffuse structure of A1351 can be a halo relic combination. Here we present the first ever spectral index map of the diffuse emission in A1351 using Giant Metrewave Radio Telescope (GMRT) 610 MHz and VLA 1.4 GHz data.
We also analyzed \textit{Chandra} X-ray data to look for a possible hint of shock front at the location of the radio-bright edge.\\
\begin{table}
\caption{A1351 Cluster properties}
\begin{tabular}{cc}
 \hline
 \hline
 Parameter & Value\\
 \hline
	RA$_{\mathrm{J2000}}$ & 11h 42m 30.8s \\
	DEC$_{\mathrm{J2000}}$ &  $+58d\ 32\arcmin\ 20\arcsec$\\
	Mass (M$_{sys}$) & $1.4-4.2\times10^{15}\ \mathrm{M}_{\odot}$\\
	Redshift ($z$) & 0.325\\
	$L_x(0.1-2.4\ \mathrm{keV})$ & 8.31$ \times 10^{44}\ \mathrm{h_{50}^{-2}\ erg\ s^{-1}}$\\
	kT & $8.69^{+1.01}_{-0.54}$ keV\\
	$\sigma_v$ & $1524^{+96}_{-74}\ \mathrm{km\ s^{-1}}$\\
 \hline
\end{tabular}
\tablecomments{Cluster properties as mentioned by \citealt{Barrena_2014MNRAS.442.2216B}}
\end{table}

The outline of the paper is as follows. We present the radio observations and data reduction procedure from GMRT and VLA in section~\ref{obs}. In Section~\ref{result} we present the results from radio observation. We discuss about the X-ray observation and results in section~\ref{x-ray} and ~\ref{xray_results} respectively. We discuss about the diffuse emission property, particle acceleration mechanism and cluster magnetic field in Section~\ref{disc}, 
followed by a summary in Section~\ref{concl}. We have assumed a $\Lambda$CDM cosmology with $\Omega_m$ = 0.3, $\Omega_{\lambda}$ = 0.7 and $H_0$ =70\,km\,s$^{-1}$\,Mpc$^{-1}$ throughout this paper. At the redshift of the Abell 1351 ($z = 0.325$), 1\arcsec corresponds to 4.704\,kpc.

\section{Radio Observation} \label{obs}
In this section, we discuss the observation and data reduction procedure of A1351 with GMRT\footnote{GMRT Data Archive: \url{https://naps.ncra.tifr.res.in/goa/data/search}} 610 MHz and VLA\footnote{VLA Data Archive: \url{https://archive.nrao.edu/archive/advquery.jsp}} 1.4 GHz.
\begin{table*}
	\centering
	\caption{The archival observation summary for GMRT and VLA and the RMS reached using uniform weighting are listed below}
	\label{tab:Table1}
	\hspace{0.01in}
    \resizebox{\textwidth}{!}{
	\begin{tabular}{cccccccc} 
	    \hline
	    \hline
		Telescope & Observation date & Frequency & Bandwidth & Time on Source & Beam & P.A. & RMS \\
		Configuration & & ( MHz) & ( MHz) & (min) & & & ($\mu$ Jy/beam) \\
	    \hline
		GMRT & Feb 2010 & 610 & 30 & 865 & $3.87\arcsec \times 3.40\arcsec $ & $+31.50$\degree & 105 \\
		VLA A & April 1994 & 1400 & 43.75 & 30 & $1.38\arcsec  \times 1.04\arcsec $ & $-3.79$\degree & 47\\
		VLA C & April 2000 & 1400 & 43.75 & 125 & $12.79\arcsec  \times 9.76\arcsec $ & $+28.43$\degree & 42\\
		VLA D & March 1995 & 1400 & 43.75 & 15 & $51.32\arcsec  \times 33.55\arcsec $ & $+49.98$\degree & 244\\
	    \hline
	\end{tabular}
	}
\end{table*}

\begin{figure*}
\begin{center}
\includegraphics[width=0.9\textwidth]{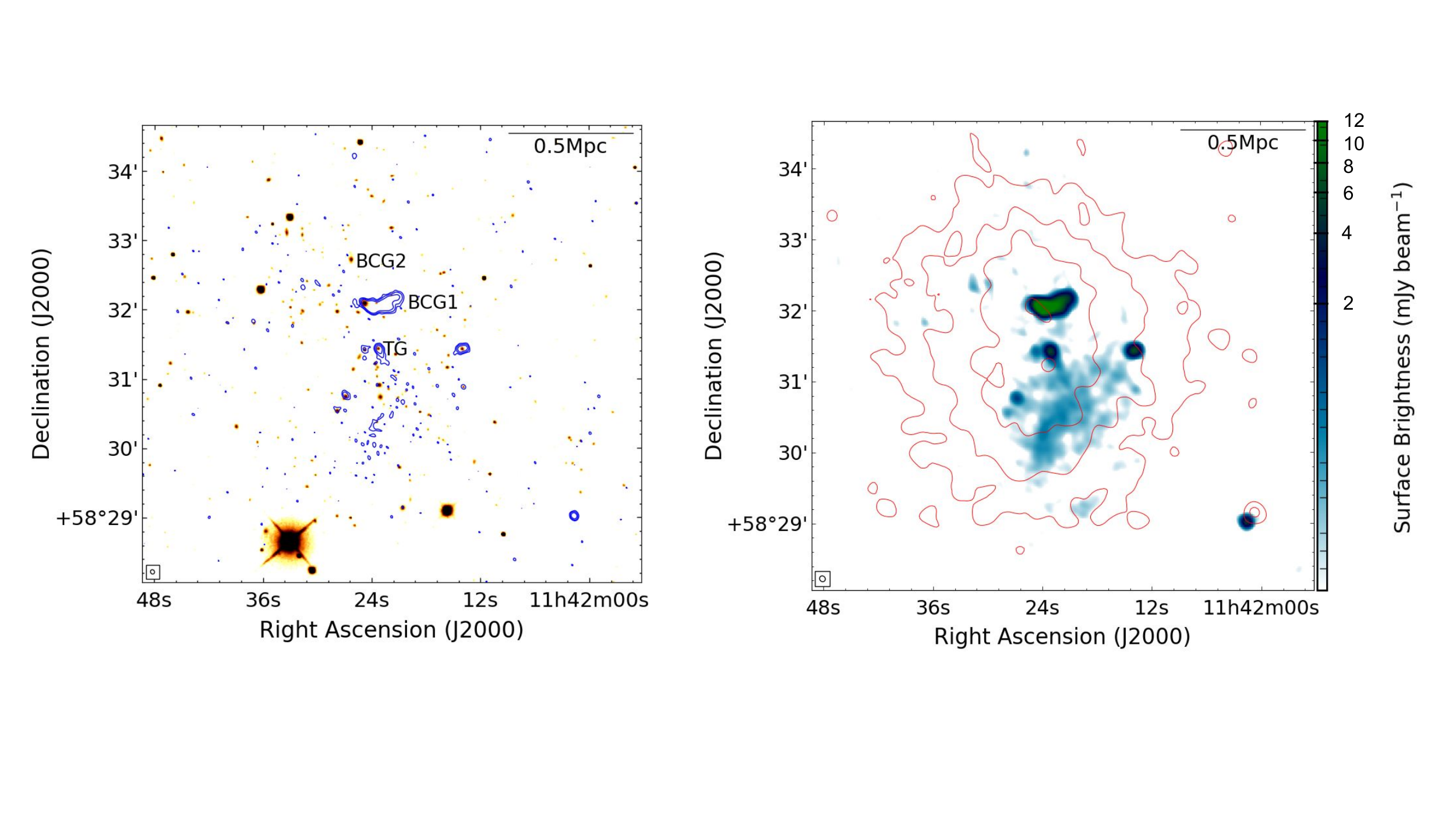}
\caption{Left: GMRT 610 MHz high resolution image contours (blue) with restoring beam $3.87\arcsec  \times 3.40\arcsec $, P.A $31.50\degree$ overlapped with SDSS-DR12 Optical image. The contours levels increases with a factor of 3 starting from 3$\sigma_\mathrm{rms}$ where $\sigma_\mathrm{rms}=105\ \mu$Jy/beam. Right: GMRT 610 MHz full resolution color image with restoring beam $5.17\arcsec \times4.67\arcsec, $ P.A 46.06$\degree$ overlaid with red X-ray contours. The X-ray contours levels increases with a factor of 2.}
\label{fig: radio_full_res}
\end{center}
\end{figure*}
\subsection{GMRT 610  MHz}
The cluster was observed with GMRT during Feb 2010 (project code \textit{17\_019} ) in dual-frequency (610 / 235 MHz) mode where the 610 MHz and 235 MHz visibilities are observed on RR and LL polarisation, respectively, with 865 min on-source time. For the GMRT Hardware Backened correlator, the bandwidth was split into upper and lower sidebands (USB and LSB), each having 16 MHz. For the two sidebands, .lta and .ltb files were generated for both frequencies. The observational summary of the cluster is presented in Table~\ref{tab:Table1}.

Source Peeling and Atmospheric Modelling (SPAM); (\citealt{refId0, Intema_refId0}), a python-based data reduction recipe, was used for the GMRT data reduction. This semi-automated pipeline uses Astronomical Image Processing System (AIPS) tasks for data reduction. 
At first, The raw data in FITS format was used for pre-calibration. In the pre-calibration part, the data from the best available scans of the primary calibrator (here 3C286) was used for calibration. \citealt{scaife_10.1111/j.1745-3933.2012.01251.x} model was used to set the flux density. For 610 MHz, data for two sidebands were pre-calibrated separately and then joined. For 235 MHz, only part of the bandwidth (USB) was available for pre-calibration. 
The pre-calibrated visibility data set of UVFITS format was used for further process. After RFI mitigation and bad data editing, several rounds of direction-independent self-calibration were performed. Finally, direction-dependent calibration was done for 610 MHz data to correct ionospheric phase errors by peeling the bright sources within the field of view. Due to the poor data quality, the direction-dependent approach could not be applied for 235 MHz, and due to the poor image sensitivity, it was not further used in our analysis. The final calibrated dataset, obtained in FITS format, was converted to Common Astronomy Software Application (CASA\footnote{\url{https://casa.nrao.edu/}}) 
measurement set (.ms) format using the CASA task \textit{importgmrt}. Further imaging was done in CASA.
\subsubsection{Imaging Diffuse Emission at 610 MHz}
\begin{figure}
\includegraphics[width=0.47\textwidth]{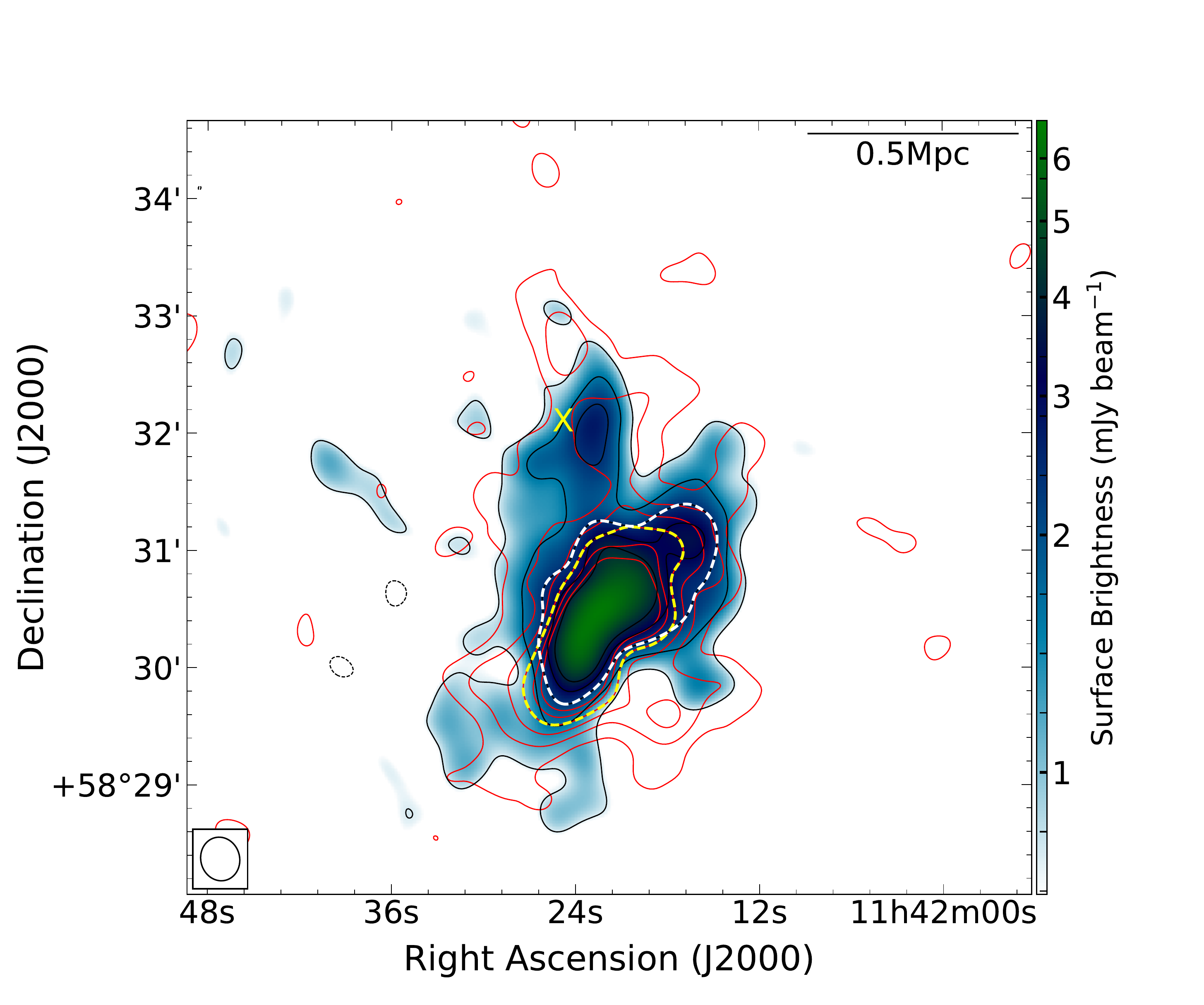}
\caption{GMRT 610 MHz low-resolution color image with restoring beam $22.46\arcsec \times 19.88\arcsec$, P.A $14.03\degree$ overlapped with VLA low resolution image contours in red and GMRT low-resolution image contours in black. The contours levels for GMRT are placed at $(-1,1,2,3,4,5,6)\times 3\sigma_\mathrm{rms}$ where $\sigma_\mathrm{rms}=260\ \mu$Jy/beam. Negative contours are dashed. The VLA contours are placed at levels $(1,2,3,4,5,6)\times 3\sigma_\mathrm{rms}$ where $\sigma_\mathrm{rms}=50\ \mu$Jy/beam. The white and yellow dashed lines shows the shape of the ridge region at 610 MHz and 1.4 GHz respectively. The yellow cross marks the location of cluster center from \citetalias{Barrena_2014MNRAS.442.2216B}.}
\label{fig: diffuse}
\end{figure}

In the full resolution ($5.17\arcsec\times4.67\arcsec$, P.A. 46.06$\degree$) image of the cluster (Fig.~\ref{fig: radio_full_res} right panel) made using Briggs \citep{Briggs_1995AAS...18711202B} robust 0, the presence of diffuse emission is visible along with the discrete radio galaxies. This bright extended emission corresponds to the halo edge or the ridge emission of A1351. The edge has a rough extension of 570 kpc at 610 MHz. To bring out the diffuse emission properly, at first, we made a high-resolution image using Briggs \citep{Briggs_1995AAS...18711202B} robust parameter -1 and selecting \textit{uv}-range greater than 1.7 k$\lambda$ (which roughly corresponds to 570 kpc at the redshift of 0.325) to ignore any contribution from the extended emission. We masked the individual radio galaxies from this image. We obtained the model visibility of these masked galaxies using CASA task \textit{tclean} on the full dataset. These model visibilities were subtracted from the data using task \textit{uvsub} in CASA. The final image for the diffuse emission was then made using Briggs robust 0, selecting a \textit{uv}-range below 20 k$\lambda$ and using 5 k$\lambda$ \textit{uv}-taper with a restoring beam $22.46\arcsec\times19.88\arcsec$, P.A. $14.03\degree$ (Figure~\ref{fig: diffuse} color map). This choice of tapering and \textit{uv}-range helped in achieving the best diffuse structure. We changed the wprojplanes parameter in \textit{tclean} to -1 to compensate for GMRT's non-coplanar baselines, which resulted in the use of 370 w-projection planes.

\subsection{VLA 1.4 GHz}
A1351 was observed with VLA A array during April 1994 (project code \textit{AB0699}) in dual polarisation mode with 30 min on-source time. It was also observed with VLA C and D configuration in dual polarisation mode on April 2000 (project code \textit{AO0149}) for 125 min and March 1995 (project code \textit{AM0469}) for 15 min respectively. The observational summary is presented in Table ~\ref{tab:Table1}. Common astronomy Software Aplication (CASA) was used for the data analysis. At first, the data for different configurations were treated individually for RFI removal and calibration. 3C286 was used as flux calibrator for all three configurations, and the flux density was set using the model described in \citet{scaife_10.1111/j.1745-3933.2012.01251.x}. After calibration was applied, the target data were split into new measurement sets for each of the data sets, and a few rounds of phase-only self-calibration were performed on the split measurement sets for all the configurations. The beam size and RMS reached using uniform weighting is given in Table ~\ref{tab:Table1}. The weights of the three data sets were equalized using CASA task \textit{statweight}, and then finally, the self-calibrated data sets were combined using CASA task \textit{concat} for further imaging.

\subsubsection{Imaging Diffuse Emission at 1.4 GHz:}
We made a high-resolution image with the combined data using Briggs robust parameter -1. To get the diffuse emission, the individual point source emissions were masked from the high resolution image and then subtracted from the visibility data as explained previously. 
Finally, after subtraction of flux contribution from individual point sources, the image for diffuse emission was made using Briggs robust 0 and selecting a \textit{uv}-range below 20 k$\lambda$ and using 5 k$\lambda$ \textit{uv}-taper (Figure~\ref{fig: diffuse} red contours).

\section{Results from Radio Observation} \label{result}
In this section, we discuss the results from GMRT 610  MHz and VLA 1.4  GHz radio Data. We also show the spectral index distribution of the diffuse emission.
\subsection{GMRT 610 MHz}
Optical study showed the cluster has two main sub-cluster in the northern region surrounding the two brightest cluster galaxies BCG1 and BCG2 (\citetalias{Barrena_2014MNRAS.442.2216B}). BCG1 also coincides with the X-ray peak. Figure~\ref{fig: radio_full_res} (left panel) is the high-resolution radio image of the cluster, made using uniform weighting, overlaid on the optical image from SDSS DATA Release 12 \citep{sdss_2015ApJS..219...12A}. BCG1 and few other galaxies are vissible in radio band, but BCG2 does not have visible radio emission. The source TG in Figure \ref{fig: radio_full_res} (left) is a tailed radio galaxy.

Figure~\ref{fig: diffuse} represents the low-resolution image of the diffuse emission after subtraction of the individual galaxies, overlapped with VLA red contours and GMRT black contours. The radio-bright edge, or the ridge (as mentioned in \citetalias{Giacintucci_2009ApJ...704L..54G} and \citetalias{Barrena_2014MNRAS.442.2216B}), has a Largest Linear Size (LLS) of $\sim$ 570 kpc with an elongation in the north-west south-east direction and a width of 260 kpc at 610 MHz. The shape of the bright ridge at 610 MHz has been highlighted with white dashed contour in Figure~\ref{fig: diffuse}. The brightest region of the edge is located $\sim$ 470 kpc away from BCG1 and $\sim$ 290 kpc away from source TG.\\

\subsubsection{Integrated Flux Density Estimation:}
For estimation of integrated Flux density of the total diffuse emission of the cluster, we selected the region within the $3\sigma_\mathrm{rms}$ contour of Figure~\ref{fig: diffuse}. The flux density within that contour region for GMRT 610 MHz was found to be $86.67  \pm 5.49$ mJy where $\sigma_\mathrm{rms}=260\ \mu$Jy/beam. For calculating the uncertainty in flux density estimation, we used the equation
\begin{equation} \label{eq1}
\Delta S=\sqrt{(\sigma_\mathrm{cal}\ S)^2+(\sigma_\mathrm{rms} \sqrt{N_\mathrm{beam}})^2} 
\end{equation}
where an uncertainty of 6\% was assumed due to calibration error($\sigma_\mathrm{cal}$) following \citealt{chandra_2004ApJ...612..974C}, ${N_\mathrm{beam}}$ is number of beams within the $3\sigma_\mathrm{rms}$ contour for GMRT image where $\sigma_\mathrm{rms}$ is the image RMS noise. The flux density of the radio edge (within 9$\sigma_\mathrm{rms}$ contour) was found 42.20 $\pm$ 2.68 mJy.\\
To crosscheck the value of flux density of the entire diffuse emission, we attempted another approach as described in \citet{Raja_2020ApJ...889..128R}. In this approach, the flux densities of the galaxies were estimated from the high-resolution image (left panel, Figure~\ref{fig: radio_full_res}) using PyBDSF (Python Blob Detector and Source Finder) \citep{Mohan_2015ascl.soft02007M}. Furthermore, these flux density values were subtracted from the total emission within the $3\sigma_\mathrm{rms}$ contour to get the amount of flux density from diffuse emission. We found the integrated flux density in this approach to be 87.18 $\pm$ 6.11 mJy.

\subsection{VLA 1.4  GHz}

\begin{figure*}
\begin{center}
\includegraphics[width=6.0in]{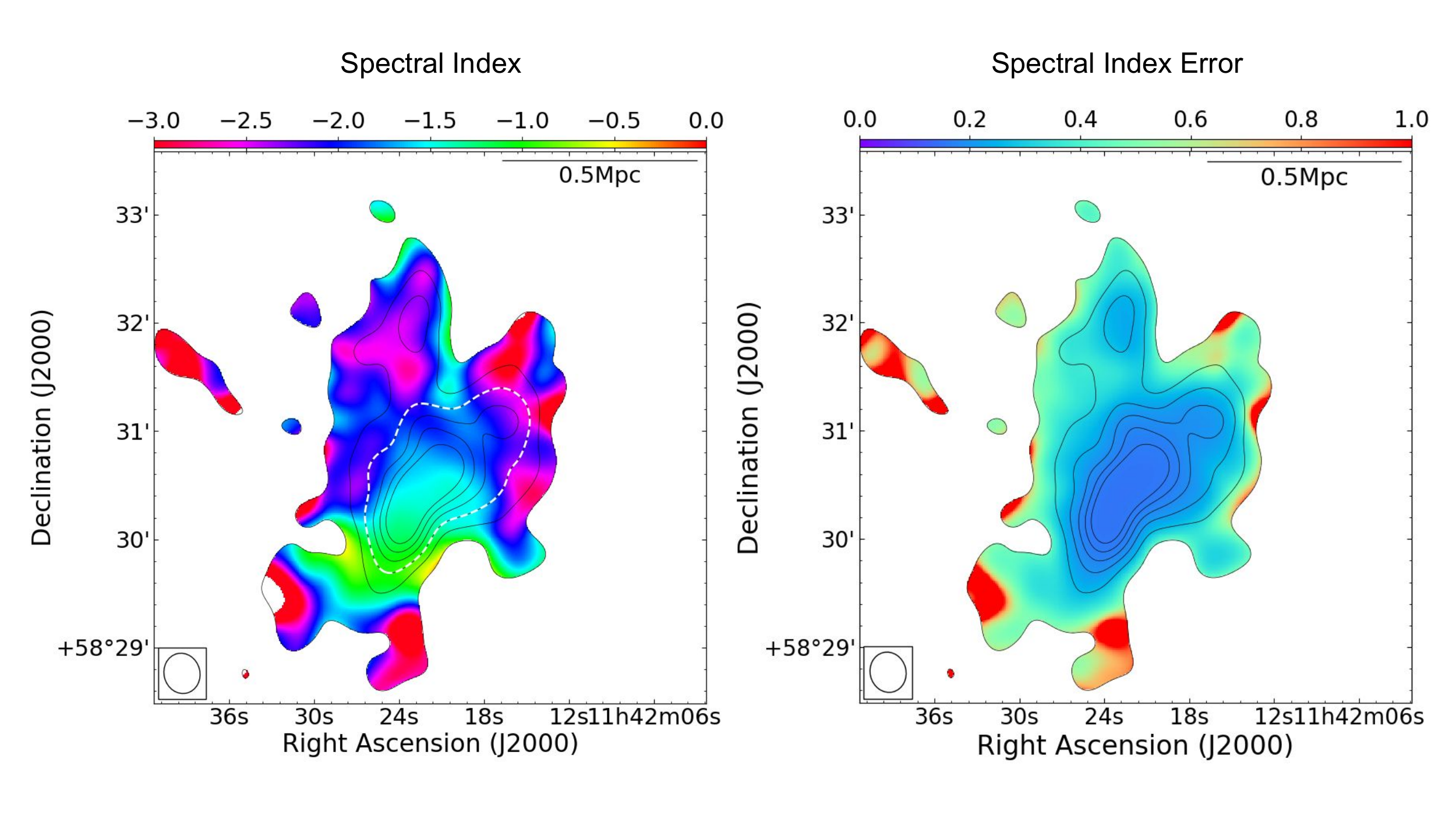} \\
\includegraphics[width=6.0in]{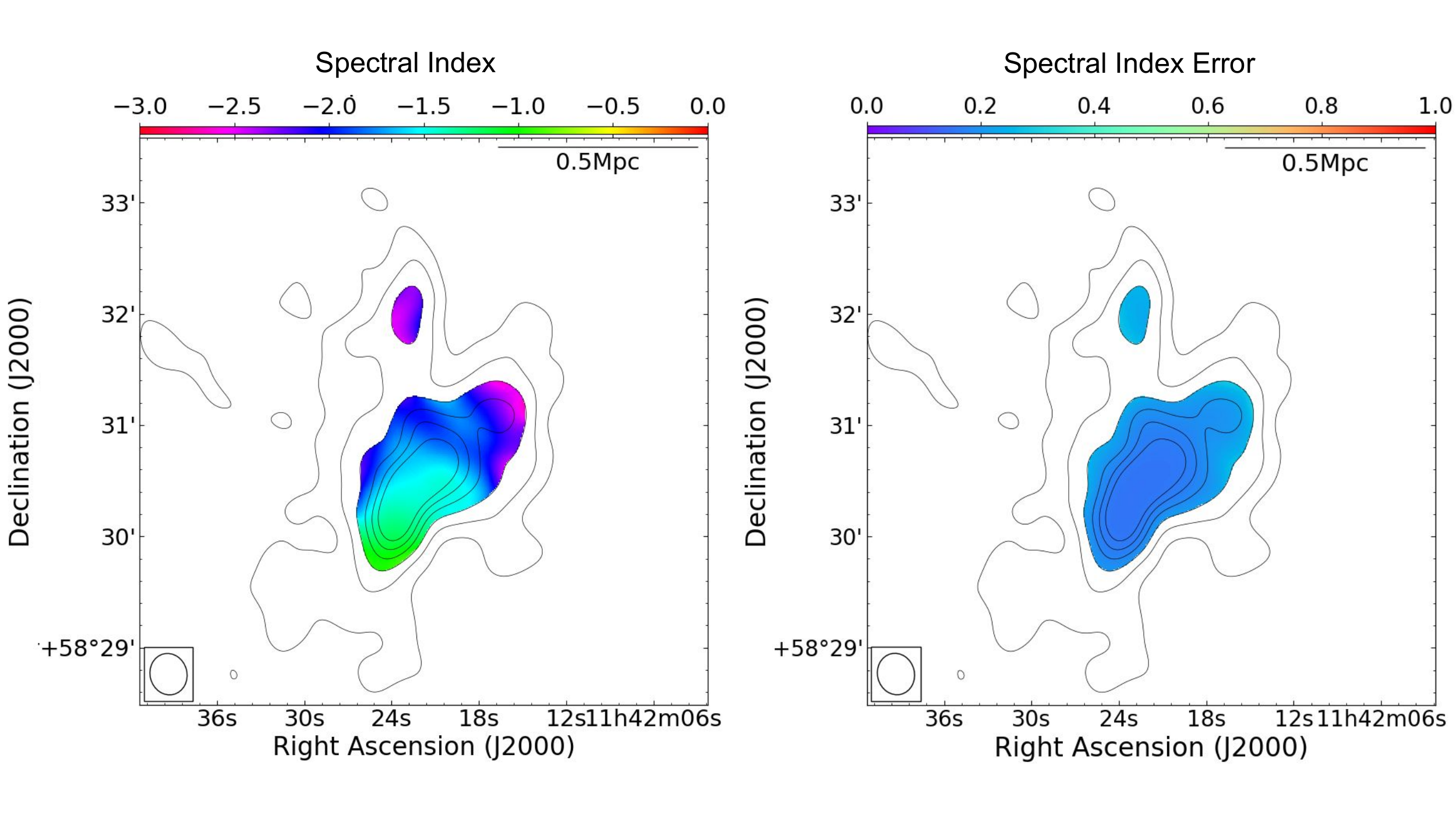} \\
\caption{Top (Left): Spectral index map between 1.4 GHz and 610 MHz with beam $22.46\arcsec \times 19.88\arcsec$, P.A. $14.03\degree$ overlapped with GMRT low-resolution image contours in Black. The contours levels for GMRT are placed at $(1,2,3,4,5,6)\times 3\sigma_\mathrm{rms}$ where $\sigma_\mathrm{rms} = 260\ \mu$Jy/beam. The edge region has been highlighted with white dashed contour. Bottom (Left): Spectral index map highlighting only the edge region. Top (Right): spectral index error map between 1.4 GHz and 610 MHz. Bottom (Right): spectral index error map for only the edge region.}
\label{fig: spec_map}
\end{center}
\end{figure*}

The red contours in Figure~\ref{fig: diffuse} represent low-resolution images obtained with VLA 1.4 GHz observation. The diffuse emission is more extented in the northern region surrounding BCG1 and BCG2 at 1.4 GHz. This emission, classified as halo previously by \citetalias{Giacintucci_2009ApJ...704L..54G}, has a quite asymmetric and elongated structure.
We have estimated the flux density of diffuse emission within $3\sigma_\mathrm{rms}$ contour to be $24.10 \pm 2.44$ mJy where $\sigma_\mathrm{rms}=50\ \mu$Jy/beam. We used the same equation (equation \ref{eq1}) for estimating error in flux density measurement assuming 10\% uncertainity due to calibration error. The flux density of the radio-bright ridge (highlighted with yellow dashed contour in Figure~\ref{fig: diffuse}) was estimated to be 10.85 $ \pm $ 1.10 mJy, which gave a radio luminosity $P_{1.4 \mathrm{\ GHz}} = 4.46 \pm 0.61\times10^{24}$ W $Hz^{-1}$. The luminosity was calculated using equation \ref{eq3}.

\begin{equation} \label{eq3}
P_{1.4 \mathrm{\ GHz}}= \frac{4\pi D_L^2(z)}{(1+z)^{(\alpha+1)}}\times S_{1.4\mathrm{\ GHz}}
\end{equation}

Where $D_L(z)$ is luminosity distance at the redshift $z$. This estimation of radio luminosity is as per the radio luminosity estimated by \citetalias{Giacintucci_2009ApJ...704L..54G} for the radio ridge region.
\begin{figure*}
\begin{center}
\begin{tabular}{lccr}
\includegraphics[width=0.5\textwidth]{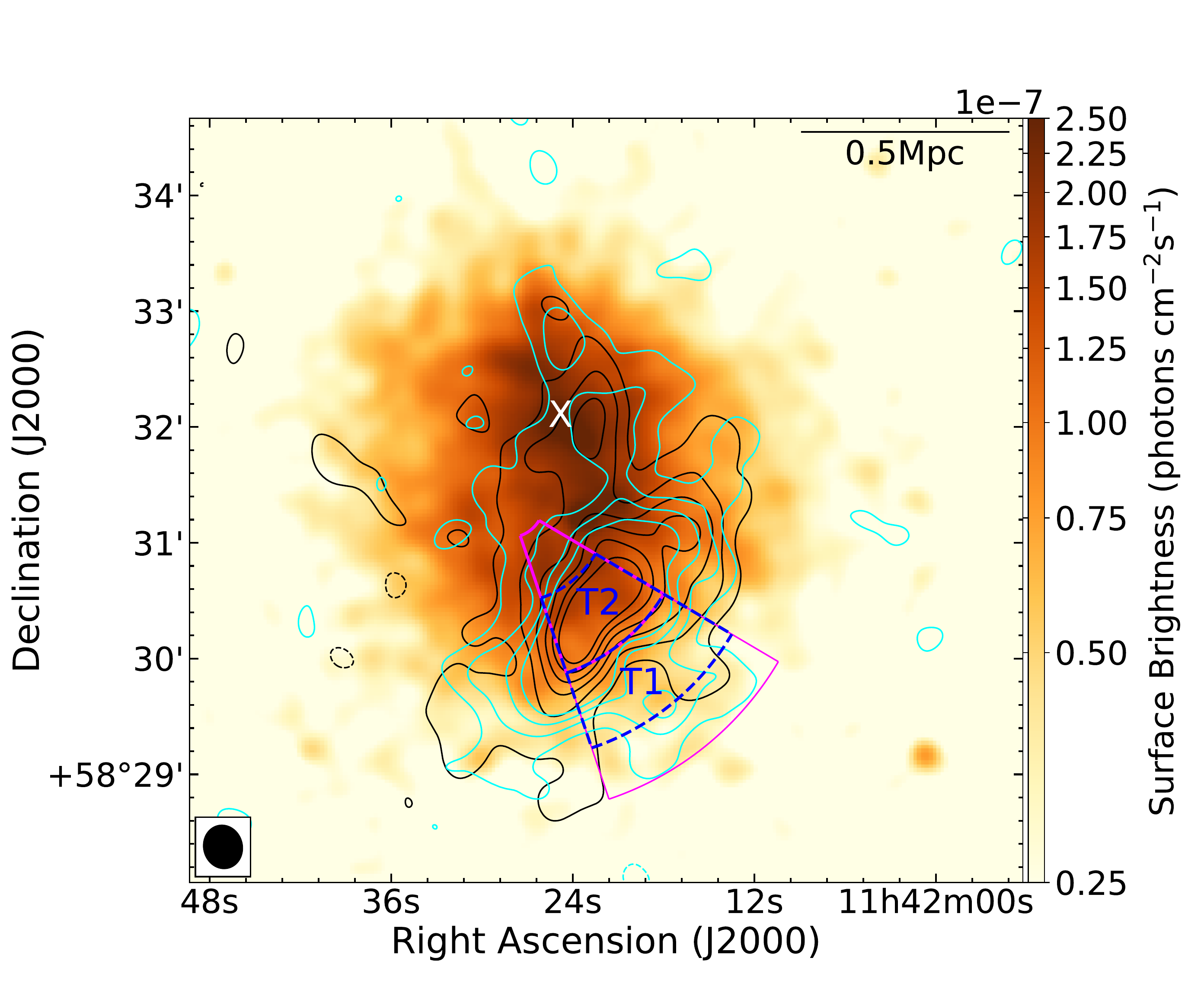} &
\includegraphics[width=0.4\textwidth]{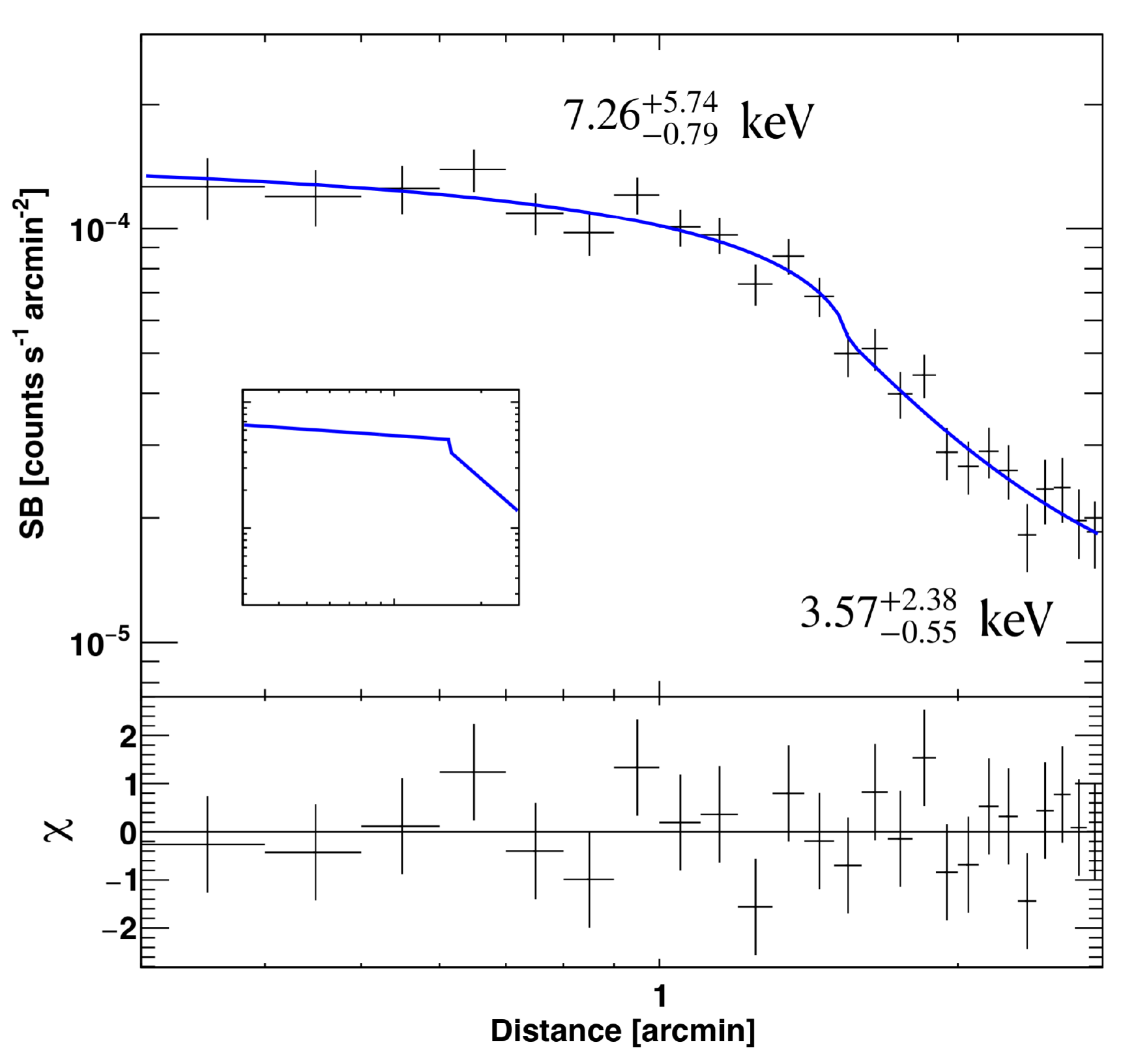} \\
\end{tabular}
\caption{Left: \textit{Chandra} X-ray surface brightness map of A1351 overlaid with
GMRT 610  MHz low resolution image contours (black) and VLA 1.4  GHz low resolution image contours (cyan) with restoring beam $22.46\arcsec \times 19.88\arcsec$, P.A. $14.03\degree$. 
The contours levels are placed at levels (-1,1,2,3,4,5,6)$\times3\sigma_\mathrm{rms}$ where $\sigma_\mathrm{rms} = 260\ \mu$Jy/beam for 610  MHz image and $\sigma_\mathrm{rms} = 50\ \mu$Jy/beam for 1.4  GHz image. The negative contours are dashed. The wedge region (magenta) was used to produce surface brightness profile, dashed regions (Blue) T1 and T2 are used to estimate the temperature across the discontinuity. The arc between the T1 and T2 region represents the position of the discontinuity. 
Right: The surface brightness profile over the wedge region (magenta, in the left panel). The inset shows the simulated gas density model. The Temperatures measured in T1 and T2 regions are $T_{1} = 3.57^{+2.38}_{-0.55}$ keV and $T_{2} = 7.26^{+5.74}_{-0.79}$ keV.}
\label{fig: xray shock}
\end{center}
\end{figure*}
\subsection{Spectral Index Estimation}

To measure the spectral index, we made images with the same \textit{uv}-range and the same restoring beam at the two frequencies. Generally, the diffuse emissions have a wider spread in the low-frequency domain. However, the northern section of the halo is not noticeable in the Figure~\ref{fig: diffuse} owing to the lesser sensitivity of the low-resolution image of 610 MHz. So, for spectral index estimation, the common region having diffuse emission present in both the frequencies was selected using CASA polygon drawing, and the flux densities in the region were noted. The spectral index was estimated using equation $\alpha=log(S_1/S_2)/log(\nu_1/\nu_2)$ where S and $\nu$ denotes flux densities at different frequencies and frequencies respectively. The integrated spectral index of the entire radio structure was found to be $\alpha_\mathrm{total} = -1.72  \pm 0.33$, where the error in the spectral index was calculated using the equation \ref{eq2}.\\
\begin{equation} \label{eq2}
\Delta\alpha=\frac{1}{\rm{log}(\nu_1/\nu_2)}\times \sqrt{\frac{(\Delta S_1)^2}{S_1^2}+\frac{(\Delta S_2)^2}{S_2^2}}
\end{equation}
We found the edge has a steep spectra with integrated spectral index $\alpha = -1.63 \pm 0.33$.
\subsubsection{Spectral Index Map:}
Spectral index mapping is an essential and informative aspect of radio data analysis as it helps better understand the particle acceleration process in the cluster. The spectral index distribution can give us an idea of where the particles are recently accelerated, thus indicating shock acceleration/re-acceleration scenarios. The spectral index of freshly accelerated particles is flatter, and it steepens as the particles lose energy post acceleration. To make the spectral index map between 610 MHz and 1.4 GHz, we made images in each frequency with Brigg's robust 0 (\citealt{Briggs_1995AAS...18711202B}), choosing the same \textit{uv}-range of 20 k$\lambda$ and applying \textit{uv}-taper of 5 k$\lambda$ in both cases. This choice of \textit{uv}-range and tapering brought out the diffuse emission best in both frequencies. The final images were obtained with the same resolution of $22.46\arcsec \times 19.88\arcsec$, PA $14.03\degree$, with the same image size and cell size. The low-frequency image was regridded using the CASA task \textit{imregrid}. The emission has less spread in some regions at 610 MHz (Figure ~\ref{fig: diffuse}). So the surface brightness below $3\sigma_\mathrm{rms}$ level of the GMRT 610  MHz images was masked. This masked region was copied to choose the same region from VLA 1.4  GHz image. Finally, the spectral index for each pixel was calculated using the CASA tool \textit{immath}. The error in the spectral index was also calculated for each pixel (Figure~\ref{fig: spec_map}).\\
Furthermore, we made the spectral index map of the bright edge/ridge region particularly, following the same way and choosing a surface brightness mask below $9\sigma_\mathrm{rms}$ level of the GMRT 610  MHz image. In Figure~\ref{fig: spec_map}, a gradient of the spectral index can be observed along the south to north direction of the edge. This gradient resembles the shock downstream steepening of the spectral index for radio relics, where when a shock passes through ICM, it accelerates or re-accelerates the electrons to relativistic speed. At the location of the shock front, as the particles are recently accelerated, the diffuse emission gives relatively flatter spectra. In the post-shock region, the high-energy particles lose energy faster through synchrotron losses and the inverse Compton effect, so a steepening in the spectral index is noticed \citep{Gennaro_2018,rajpurohit2020A&A...636A..30R}. 

\subsection{Radio Mach Number}

In the DSA model of particle acceleration, the particles from the thermal pool of ICM travel back and forth in the pre and post-shock region and gain energy in this process \citep{Hoeft_2007MNRAS.375...77H, Brunetti_2014IJMPD..2330007B}. This model predicts that the shock Mach number for planar shock $\mathcal{M}$ is related to the integrated spectral index over a region with different plasma ages as
\begin{equation}
  \mathcal{M}_{\alpha} = \sqrt{\frac{(\alpha-1)}{(\alpha+1)}} 
\end{equation}
For the bright ridge/edge in A1351 we found a shock Mach number of 2.05 considering the integrated spectral index $\alpha$ = -1.63 over the region that shows spectral steepening similar to radio relics (Fig \ref{fig: spec_map} Bottom Left).
\section{Chandra X-ray Observation}
\label{x-ray}
We analyzed archival \textit{Chandra}\footnote{\textit{Chandra} Data Archive: \url{https://cda.harvard.edu/chaser/}} X-ray observations (ObsId 15136) of A1351. The total 33 ksec observations were taken in VFAINT mode.
For this study, we employed a systematic calibration and analysis pipeline, which uses the \textit{Chandra} Interactive Analysis of Observations (CIAO) and subsequent scripts in IDL and python.
The details of our data reduction pipeline are described in \citep{Datta_2014ApJ...793...80D, Schenck_2014AJ....148...23S, Hallman_2018ApJ...859...44H, Raja_2020ApJ...889..128R, Rahaman_2021MNRAS.505..480R}, which initially consisted of several bash and IDL scripts. 
As A1351 is a comparatively low redshift cluster ($z = 0.325$), the extended X-ray emission spills over all four ACIS-I chips. Therefore, we were not able to use the local background (as in, e.g., \citealt{Raja_2020ApJ...889..128R}). Hence, we subtracted the background contribution present in the observation by extracting background spectra from the "blank-sky" background files. These `blank-sky' background files are available in the \textit{Chandra} calibration database (CALDB), which represents particle background and unresolved cosmic X-ray background. The ``blank-sky'' background was re-normalized in the 9.5 - 12 keV band. The \textit{Chandra} effective area is negligible in the 9.5 - 12 keV band, and all the 9.5 - 12 keV flux in the sky data is due to particle background \citep{Hickox2006ApJ...645...95H}.

\begin{table*}
	\centering
	\caption{Best fit parameters of broken power law model (using PROFFIT) are listed below.}
	\label{tab:Table2}
	\resizebox{\textwidth}{!}{
	\begin{tabular}{lcccccc} 
		\hline
		\hline
		$\alpha1$ & $\alpha2$ & $r_\mathrm{sh}$(arcmin) & norm & Jump ($C$) & $\chi^2/D.o.f$ & $\mathcal{M}_\mathrm{SB}$\\
		\hline
		0.16 $\pm$ 0.14 & 2.01 $\pm$ 0.35 & 1.55 $\pm$ 0.01 & $5.43 \pm  2.02 e-05$ & 1.24 $\pm$ 0.25 & 15.80/18 (0.88) & $1.34^{+0.20}_{-0.19}$ \\
		\hline
	\end{tabular}
	}
\end{table*}
Next, we removed point sources from the data by providing SAOImage DS9 region file containing point sources. These point sources were detected using the tool {\it wavdetect} inbuilt into CIAO in the 0.7 - 8.0 keV band with the scales of 1, 2, 4, 8, and 16 pixels. These were further inspected visually for any false detection or if {\it wavdetect} failed to detect any real point sources. Regions with point sources were removed from both data and blank-sky background files to avoid negative subtraction. 
After removing the point sources, we created light curves for individual ObsId in full energy and 9.5 - 12 keV bands. Light curves were binned at 259 seconds per bin for data as well as blank-sky backgrounds, and count rates higher than 3$\sigma$ were removed (background flares) using the \textit{deflare} tool.
These steps produced calibrated and clean data free from bad events as well as contaminating point sources.

After cleaning the data, we used \textit{merge\_obs} with binning 4 to produce a surface brightness map. The exposure corrected, background and point sources subtracted, 0.7 - 8.0 keV surface brightness image is shown in Figure \ref{fig: xray shock}.

We performed spectral analysis using XSPEC version: 12.9.1 \citep{Arnaud_1996ASPC..101...17A} in between 0.7 - 8.0 keV energy range. 
We used  \textit{CIAO} task \textit{dmextract} to create spectra from each region from each observations and \textit{specextract} to calculate  Auxiliary Response File (ARF) and Redistribution Matrix File (RMF). 
The APEC (Astrophysical Plasma Emission Code) and PHABS (PHotoelectric ABsorption) models were fitted to the spectra from each region using C-statistics \citep{Cash1979}. The metallicity of the cluster was kept frozen at 0.3 $Z_\odot$ throughout the cluster, where $Z_\odot$ is the solar abundance \citep{Anders_1989GeCoA..53..197A}. Redshift (0.325; \citetalias{Barrena_2014MNRAS.442.2216B}) and $N_\mathrm{H}$ ($7.12 \times 10^{19}\mathrm{cm^{-2}}$; \citealt{HI4PI_2016A&A...594A.116H}) were also kept frozen. Only APEC normalization and temperature were fitted for each region.

\section{Results from X-ray Observation}\label{xray_results}

Figure~\ref{fig: xray shock} shows the X-ray emission from the cluster is slightly elongated towards the N-NE S-SW direction, which implies a merging cluster \citetalias{Barrena_2014MNRAS.442.2216B}. The cluster also hosts a large-scale diffuse radio emission (a radio halo and a radio bright edge/ridge). The contour of the radio edge is denser towards the southwest direction. These denser contours may have resulted due to compression by a shock front. Therefore, to look for a possible hint of cluster shock, we analyzed the cluster X-ray surface brightness profiles using PROFFIT v1.5 \citep{Eckert_2011A&A...526A..79E, Eckert_2016ascl.soft08011E}. 

\subsection{X-ray Surface Brightness Discontinuity:}
To estimate any discontinuity in surface brightness profile, we proceeded in the following way. We created a number of concentric annuli over the wedge region with $6\arcsec$ bin size.
This bin size was chosen to get a sufficient count in each bin.
We iteratively choose the wedge region to maximize the jump of the surface brightness discontinuity.
The surface brightness profile over the wedge region was fitted with a broken power-law model (bknpow, in-build in PROFFIT v1.5). The broken power-law density model can be defined as

\begin{equation}
\begin{split}
n(r) & = Cn_0\Big(\frac{r < r_\mathrm{sh}}{r_\mathrm{sh}}\Big)^{-\alpha1}\\
    &  = n_0\Big(\frac{r > r_\mathrm{sh}}{r_\mathrm{sh}}\Big)^{-\alpha2}
\end{split}
\end{equation}

where $C$ is the compression factor, $n(r)$ is electron number density, $n_0$ is normalization constant, $r_\mathrm{sh}$ is the radial distance of the shock, $r$ is the distance from the center of the wedge region, and $\alpha1$ and $\alpha2$ are the power-law indices for the respective profiles.
We found a discontinuity in the surface brightness profile over the wedge region (magenta) shown in Figure \ref{fig: xray shock}.
The best-fitted parameters and the profile is shown in Table~\ref{tab:Table2} and Figure~\ref{fig: xray shock} (right panel) respectively. 
The position of the discontinuity in surface brightness is shown with an arc between the T1 and T2 regions in the left panel of Figure \ref{fig: xray shock}.
The broken power law was fitted with a reduced $\chi^2$ value of 0.88. 

\subsection{X-ray Temperature Jump:}
To characterize the surface brightness discontinuity as shock or cold front, we estimated the temperature across the regions of the discontinuity. As the exposure of the \textit{Chandra} observation was short, we were not able to make a temperature profile, but from the dedicated regions across the discontinuity (e.g., T1 and T2 in Figure \ref{fig: xray shock}).
We iteratively choose the width of the region to have threshold counts in the corresponding spectra. We took regions with 41$\arcsec$ of width, labeled as T1 and T2 in the left panel of Figure \ref{fig: xray shock}. The temperature was estimated by fitting the thermal APEC and PHABS model as discussed in section \ref{x-ray}.
The estimated temperatures in the pre-shock or upstream region, $T_{1}$ and post-shock or downstream
region, $T_{2}$ are listed in Table \ref{tab:xray_temp}, where $T_{1}$ and $T_{2}$ are temperatures from the regions T1 and T2 respectively as labeled in Figure \ref{fig: xray shock} (left panel). 
We found that there is a significant jump in the temperature along the direction similar to the surface brightness discontinuity. Therefore, the presence of both discontinuities hints towards the shock front.
\begin{center}
\begin{table}[h!]
\caption{\label{tab:xray_temp}Shock upstream (T$_{1}$) and downstream T$_{2}$ temperatures and Mach number from temperature jump is listed below.}
\begin{tabular}{ l c r }
 \hline
 \hline
 T$_{1}$(T$_{up}$) & T$_{2}$(T$_{down}$) & $\mathcal{M}_\mathrm{T}$ \\ 
 \hline
 $3.57^{+2.38}_{-0.55}$ keV & $7.26^{+5.74}_{-0.79}$ keV & $1.96^{+1.36}_{-0.86}$\\ 
 \hline
\end{tabular}
\end{table}
\end{center}

\subsection{X-ray Mach Numbers}
Assuming the jumps as shock front, we calculated the Mach number using the Rankine-Hugoniot shock jump condition given as
\begin{equation}
    \frac{\rho_2}{\rho_1} = C = \frac{(1+\gamma)\times \mathcal{M}_\mathrm{SB}^2}{2+(\gamma - 1)\times \mathcal{M}_\mathrm{SB}^2}
\end{equation}
where $\rho_1$ \& $\rho_2$ are densities at pre and post shock region respectively and for mono-atomic gas, considering $\gamma = 5/3$, we get 
\begin{equation}
    C = \frac{4\mathcal{M}_\mathrm{SB}^2}{3+\mathcal{M}_\mathrm{SB}^2}
\end{equation}
This gives a shock Mach number from density jump across the shock edge, $\mathcal{M}_{SB} = 1.34^{+0.20}_{-0.19}$. \\
The Mach number from temperature jump was calculated using Rankine-Hugoniot temperature jump condition
\begin{equation}
    \frac{T_2}{T_1} = \frac{(5\mathcal{M}_\mathrm{T}^2 - 1)(\mathcal{M}_\mathrm{T}^2 + 3)}{16\mathcal{M}_\mathrm{T}^2}
\end{equation}
where, $T_1$ and $T_2$ are the pre-shock (upstream) and post-shock (downstream) temperature respectively (Table \ref{tab:xray_temp}). 
The Mach number from temperature jump was found to be $\mathcal{M}_\mathrm{T} = 1.96^{+1.36}_{-0.86}$.

\section{Discussion} \label{disc}
\begin{figure}
\includegraphics[width=0.49\textwidth]{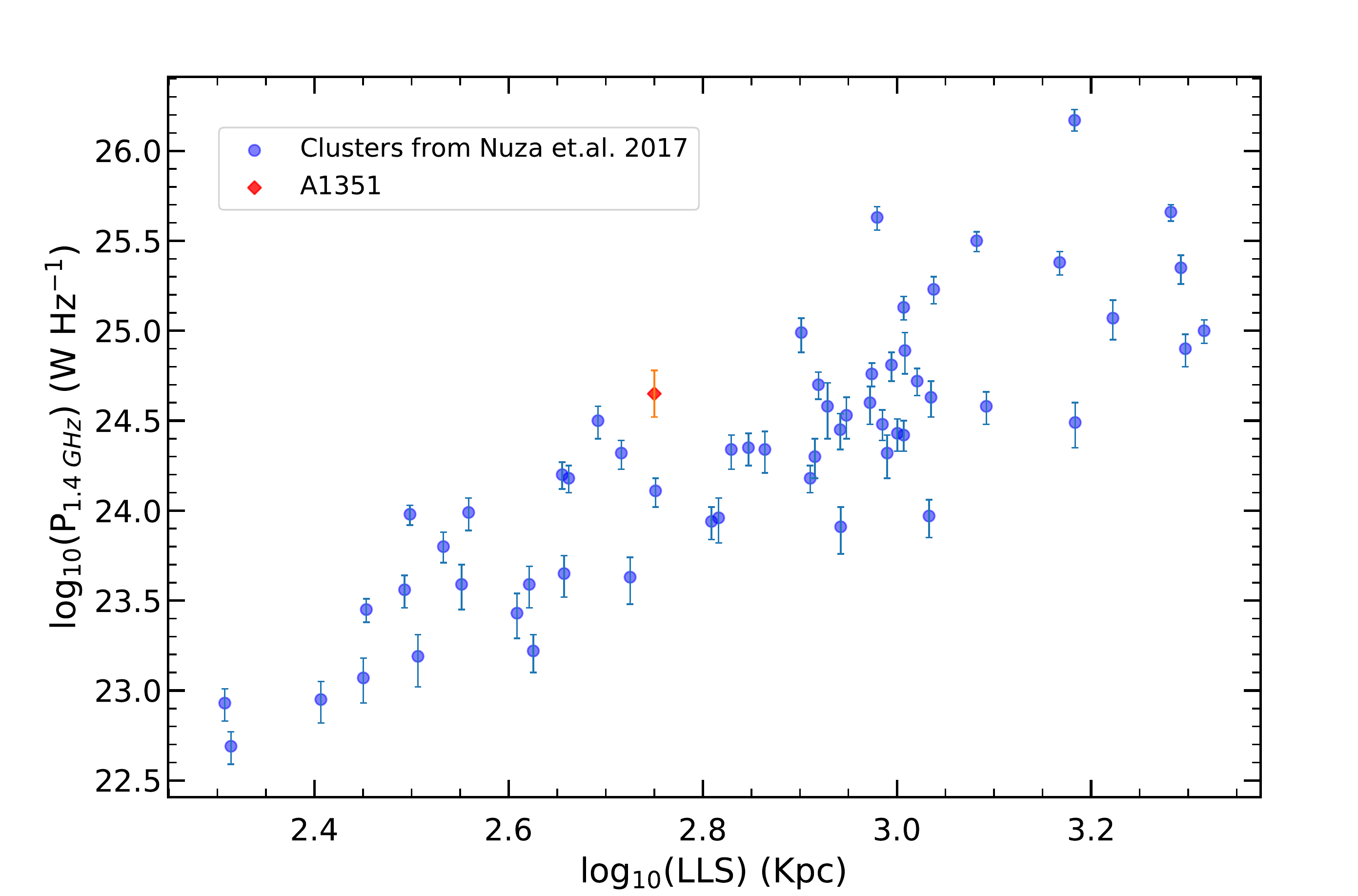}
\caption{Radio Luminosity $P_{1.4\ \mathrm{GHz}}$ vs. Largest Linear Size (LLS) of the relics adopted from the literature (\citealt{Nuza_2017MNRAS.470..240N}). The relic in A1351, marked with a red diamond, follows observed trend and given its size, it is one of the highly luminous relics. }
\label{fig: lls_P}
\end{figure}
The diffuse radio emission in A1351 is quite asymmetric in nature with the presence of the bright region at the southern part of the radio halo (Figure \ref{fig: diffuse}). This type of radio-bright edges within radio halos has been noticed previously for a few clusters but with smaller sizes (e.g. \citealt{Markevitch2005, Macario_2011, Shimwell_2014MNRAS.440.2901S,Wang_2018}). Here, the bright edge has an LLS of $\sim$570 kpc at 610 MHz and situated at a distance of $\sim$470 kpc away from the cluster center. This size and location indicate that the edge can be a relic. As mentioned in \citetalias{Barrena_2014MNRAS.442.2216B}, the cluster is undergoing a merging in the N-NE S-SW direction where the passage of an axial shock can form the relic. The spectral index distribution in A1351 is in agreement with the other reported cases where cluster radio shocks resulted in radio relics in cluster peripheral region (see review by \citealt{vanweeren_2019SSRv..215...16V}). We have compared the LLS and luminosity ($P_{1.4\ \mathrm{GHz}}$) of the relic in A1351 with other known relics from \citealt{Nuza_2017MNRAS.470..240N} (Figure \ref{fig: lls_P}). We see that the relic in A1351 (Marked in Red diamond) follows the observed trend in $P_{1.4\ \mathrm{GHz}}$-LLS plot. Given its LLS, The relic is one of the highly luminous relics with a radio power $P_{1.4\ \mathrm{GHz}}$ = $4.46\times10^{24}$ W $Hz^{-1}$.

For the case of A1351, the shock Mach number derived from Radio observation ($\mathcal{M}_{\alpha}$ = 2.05) is slightly higher than that observed from the X-ray temperature jump ($\mathcal{M}_T$ = $1.96^{+1.36}_{-0.86}$) and density jump ($\mathcal{M}_{SB}$ = $1.34^{+0.20}_{-0.19}$). The discrepancy is expected as the radio synchrotron emission is dependent on the amplitude of the magnetic field fluctuations. These fluctuations decrease slower than the density and temperature fluctuations leading to the higher values of radio-derived Mach number seen for most radio relics \citep{Domnguez_Fernndez_2020}. \citet{Wittor_2021MNRAS.506..396W} showed the radio-derived Mach numbers are more skewed to the higher value of Mach number distribution while the X-ray derived Mach numbers reflect their average distribution. Moreover, the discrepancy observed in the derived values of Mach numbers is due to the systematic errors that can affect the different analyses. 
The radio estimations can suffer errors in flux measurement due to shallow data or source contamination \citep{Hoang_2017MNRAS}. Also, the measurement of Mach number from spectral index is challenging. Mach number measured using the injection index close to the shock front is theoretically more precise in determining the shock properties. However, it is more difficult to achieve as it needs identification of the position of the shock precisely. An experimentally more robust approach to Mach number is using the volume-averaged spectral index. Although, this approach is theoretically less precise as it involves electrons from different ages across the relic \citep{Colafrancesco_2017MNRAS, Wittor_2021MNRAS.506..396W}.
On the other hand, the influence of the projection effect on X-ray analyses makes the radio findings a more robust technique \citep{Hong_2015ApJ,Akamatsu_2017A&A}. 
In case of temperature estimation, there is the possibility of both underestimating the post-shock temperature as well as overestimating the pre-shock temperature, which can lead to underestimation of Mach number ($\mathcal{M}_T$). However, the temperature jump is still less subject to the projection effect than the density jump and thus the former one gives a more reliable result \citep{Ogrean_2013MNRAS, Akamatsu_2017A&A}.

In our case, the value of the shock Mach number obtained from radio observation is quite consistent with the shock Mach number derived from the X-ray temperature jump.
This supports the DSA mechanism of particle acceleration at the shock front. The spectral index gradient is also in accordance with that. There have been very few radio relics in the literature where DSA mechanism for particle acceleration with weak shocks could be established \citep{Bourdin_2013ApJ...764...82B, shimwell_2015MNRAS.449.1486S, botteon2016elgordoMNRAS.463.1534B, Locatelli_2020MNRAS.496L..48L, Rajpurohit_2021arXiv210405690R}. The relic in A1351 is a viable candidate for supporting DSA.

\subsection{Electron acceleration efficiency}\label{acc_eff}
The LLS - P$_{1.4\ \mathrm{GHz}}$ graph (Figure \ref{fig: lls_P}) shows that the relic in A1351 is highly luminous, which would require high acceleration efficiency for the thermal electrons. To check the consistency of the DSA model and the electron acceleration efficiency needed to produce the observed radio luminosity for A1351, we used the relationship as described in \citealt{Hoeft_2008MNRAS.391.1511H,Locatelli_2020MNRAS.496L..48L}. The acceleration efficiency $\zeta_e$ is related to the cluster magnetic field B, X-ray properties, and radio luminosity $P_\nu$ at frequency $\nu$ by the equation

\begin{multline}
\zeta_e = \frac{P_\nu}{6.4\times10^{-34}} \frac{B^2+B_\mathrm{CMB}^2}{B^{1-\frac{\alpha}{2}}} \frac{\mathrm{Mpc^2}}{A}\Big(\frac{\nu}{1.4\ \mathrm{GHz}}\Big)^{\frac{-\alpha}{2}}\\
\Big(\frac{7\ \mathrm{keV}}{T_\mathrm{d}}\Big)^{\frac{3}{2}} \frac{10^{-4}}{n_\mathrm{ed}}   
\end{multline}

\begin{figure}
\includegraphics[width=0.47\textwidth]{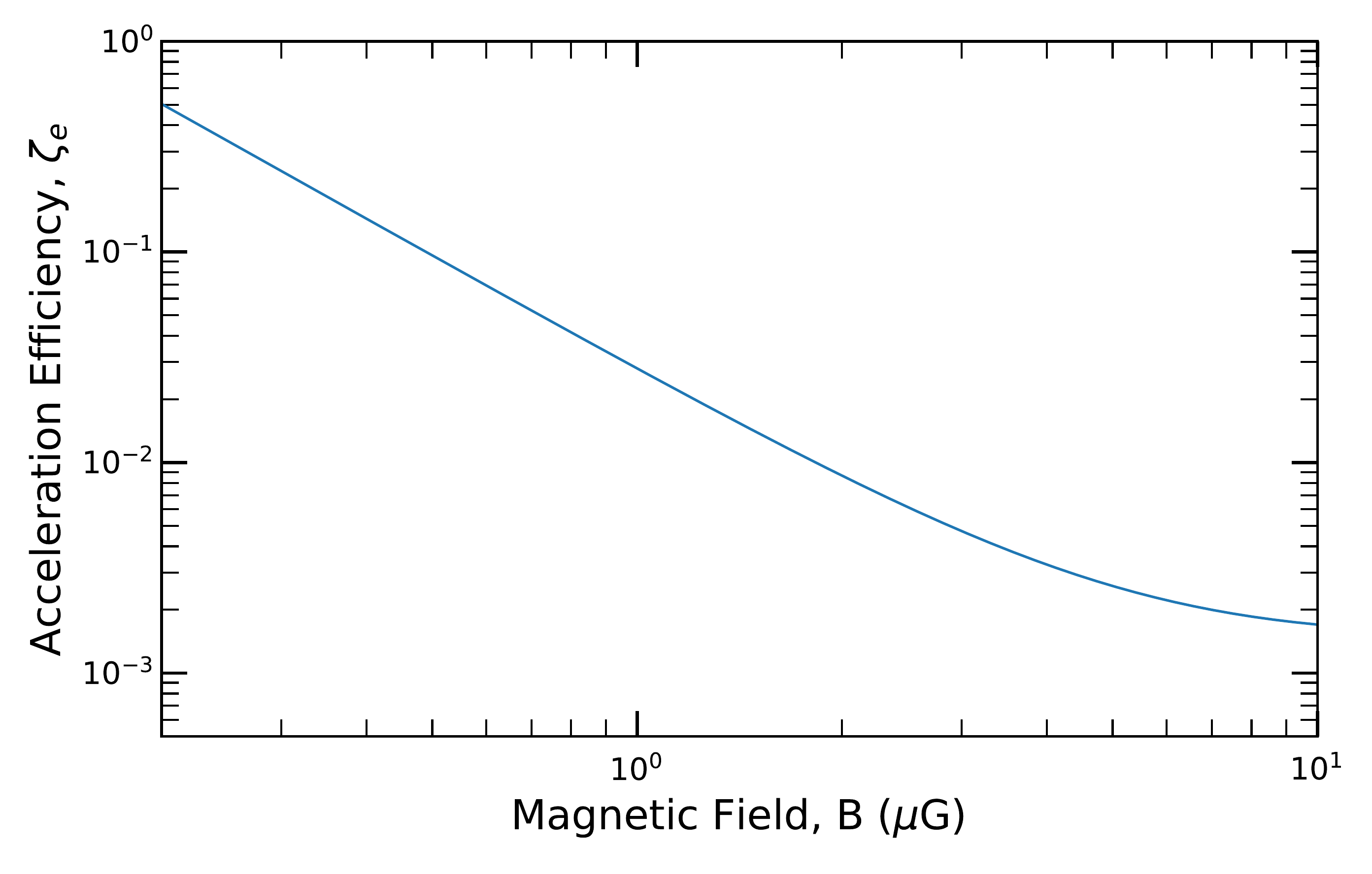}
\caption{The electron acceleration efficiency $\zeta_e$ plotted against magnetic field at the shock location. The value of the magnetic field has been kept within few $\mu$G.}
\label{fig: xi}
\end{figure}

Where $\alpha$ is the spectral index -1.63, $T_\mathrm{d}$ is the downstream temperature $7.26^{+5.74}_{-0.79}$ keV, $n_\mathrm{ed}$ is the downstream electron density [$0.26^{+0.12}_{-0.15}\ \times\ 10^{-3}$] cm$^{-3}$ and $B_{\rm CMB}$ is the equivalent magnetic field of Cosmic Microwave Background at the cluster redshift. The surface area, $A$ of the relic was taken as LLS $\times$ LLS following \citep{Locatelli_2020MNRAS.496L..48L,Rajpurohit_2020A&A...636A..30R}. $B_{\rm CMB}$ was evaluated at redshift 0.325 using the equation \ref{eq9} following \citealt{Hoeft_2007MNRAS.375...77H}.

\begin{equation}
\label{eq9}
B_\mathrm{CMB} = 3.24(1+z)^2 \mu G
\end{equation}

Figure \ref{fig: xi} shows that in the presence of a few $\mu$G magnetic fields (0.5 - 5 $\mu$G), the efficiency $\zeta_e$ at the shock location varies from $\sim10\%$ to $\sim0.2\%$. This range of acceleration efficiency is comparable to other cases where DSA model could explain the observed luminosity of the respective radio relics (e.g. A521, El Gordo \citealt{Botteon2020A&A...634A..64B}). The high luminosity and the weak shock with $\mathcal{M} \sim 2.05 $ raise the possibility that DSA might not be the sole mechanism for powering the relic and that shock re-acceleration might be involved as well. However, in presence of a strong magnetic field, the acceleration of thermal electrons via the DSA is still a possible option for the relic in A1351. Deeper observations are required to shed more light about the particle acceleration process.

\subsection{Equipartition Magnetic Field}\label{B_eq}

Cluster magnetic field can be approximated assuming equipartition of energy between cosmic ray particle and magnetic field (\citealt{Govoni_2004,Bonafede_2009,vanWeeren_2009,Parekh_2020MNRAS.499..404P,Pandge2022}). To quantify the magnetic field at the relic location in A1351, we assumed that the system satisfies minimum energy density condition. The minimum energy density $u_\mathrm{min}$ is obtained using equation \ref{eq11} following \citealt{Govoni_2004}.

\begin{equation} \label{eq11}
u_\mathrm{min} = \xi(-\alpha,\nu_1, \nu_2)(1+k)^{\frac{4}{7}}(\nu_0)^{\frac{-4\alpha}{7}}(1+z)^{\frac{12-4\alpha}{7}}
\Big(\frac{I_0}{d}\Big)^{\frac{4}{7}}
\end{equation}

Where $\xi(-\alpha,\nu_1, \nu_2) = 1.57 \times 10^{-13}$ for a frequency range $\nu_1$(10 MHz) - $\nu_2$(100 GHz) and spectral index $\alpha$ = -1.63 \citep{Govoni_2004}, $\nu_{0}$ = 1400 MHz, $I_0$ was estimated by dividing the flux density (10.82 mJy) of the relic at 1.4 GHz by the solid angle of $85\arcsec \times 103\arcsec$, and the depth $d$ of the relic was obtained in kpc taking an average of the largest (550 kpc at 1.4 GHz) and smallest (220 kpc at 1.4 GHz) linear size of the relic at 1.4 GHz. We assumed the ratio of the energy content of cosmic-ray protons and electrons $k = 1$ \citep{Parekh_2020MNRAS.499..404P}.
Thereafter, the equipartition magnetic field $B_\mathrm{eq}$ can be found using equation \ref{eq12}
\begin{equation} \label{eq12}
    B_\mathrm{eq} = \Big(\frac{24\pi}{7} u_\mathrm{min} \Big)^{1/2}
\end{equation}
The magnetic field found with this approach at A1351's relic location is $\sim$ 1.8 $\mu$G.

A modified formula to compute the magnetic field strength of synchrotron radio sources ($B\arcmin_\mathrm{eq}$) is to consider an upper and lower energy cut-offs for the cosmic ray electrons rather than frequency cut-offs \citep{Brunetti_1997, Beck_2005AN....326..414B}. Assuming  $\gamma_\mathrm{min} = 100\ (<< \gamma_\mathrm{max})$, the revised magnetic field is calculated using equation \ref{eq13}.
\begin{equation}\label{eq13}
    B\arcmin_\mathrm{eq} \sim 1.1 \gamma_\mathrm{min}^{\frac{1+2\alpha}{3-\alpha}} B_\mathrm{eq}^\frac{7}{2(3-\alpha)}
\end{equation} 
Where $\gamma$ is the Lorentz factor, $B_\mathrm{eq}$ is the equipartion magnetic field obtained using equation \ref{eq12}. The revised magnetic field found for A1351 relic is $\sim 5.3\ \mu$G. The equipartition theory relies on several assumptions. The particle energy distribution at the cluster is poorly known. Projection effects can influence the extent of the relic, and the assumption of the relic's depth can affect the derived value of magnetic field. A better constraint on the magnetic field can be given through Faraday Rotation Measure (RM). For few radio relics, the magnetic field derived using RM study are in agreement with equipartition estimation  (eg. coma cluster: \citealt{Bonafede2013}, MACSJ0717.5$\pm$3745: \citealt{Bonafede2009macs,Rajpurohit2022} ). However, for few cases the magnetic fields derived from these two methods differ (eg. A2345: \citealt{Bonafede2009_A2345,Stuardi2021}; A3667: \citealt{Johnston-Hollitt2003PhD,deGasperin2021}). Polarisation study is needed to give a better estimation of the magnetic field.

\section{Summary}
\label{concl}
In this paper, we present the first time analysis and result of GMRT 610 MHz, Chandra X-ray data of A1351 and a reanalysis of the VLA 1.4  GHz data to understand the origin of the peculiar radio edge emission present in the cluster. With \textit{Chandra} data we searched for any possible hint of shock at the location of the bright radio edge in A1351. Our findings are summerized below.
\begin{itemize}
    \item From radio observation, we measured a total diffuse emission flux at 610  MHz and 1.4  GHz to be $86.67  \pm 5.49$ mJy and $24.10  \pm 2.44$ mJy, respectively. The average spectral index of the cluster diffuse emission was found to be $\alpha_\mathrm{total} = -1.72  \pm 0.33$. The bright edge of the cluster has an LLS of $\sim$ 570 kpc at 610 MHz, radio luminosity $P_{1.4\ \mathrm{GHz}}$ = $4.46\times10^{24}$ W $\mathrm{Hz}^{-1}$ and an integrated spectral index $\alpha = -1.63 \pm 0.33$ giving a Mach number $\mathcal{M}_{\alpha}$ = 2.05. The spectral index map shows a spectral index gradient at the edge location.
    \item Using \textit{Chandra} observation, we found the presence of shock at the location of the edge. The shock Mach numbers found from X-ray temperature and density jumps are $1.96^{+1.36}_{-0.86}$ and $1.34^{+0.20}_{-0.19}$, respectively.
    \item Its larger size, gradient in the spectral index map, and position of the X-ray shock support the idea that the edge is a radio relic. The relic follows the $P_{1.4\ \mathrm{GHz}}$-LLS trend with other observed relics from the NVSS survey and stands as one of the powerful relics considering its size.
    \item We found a magnetic field of $\sim 5.3\ \mu$G in the relic's location assuming equipartition. In presence of this high magnetic field, it is very likely that the relic in A1351 might have originated from DSA of thermal electrons. The formation of this highly luminous relic in presence of weak shock makes it one of the unique cases which can unfold interesting information about cluster scale acceleration processes.
    \item Future deep X-ray observation can help us study the cluster X-ray morphology in more detail. Future polarisation study can give better understanding of the cluster magnetic field. Multi-frequency radio observations are needed to search for spectral curvature and also to study this halo relic system in more detail. 
    
\end{itemize}

\section*{acknowledgements}
We thank the anonymous reviewer for the comments and suggestions. We thank IIT Indore for giving out the opportunity to carry out the research project. MR would like to thank DST for INSPIRE fellowship program for financial support (IF160343). MR also acknowledges financial support from Ministry of Science and Technology of Taiwan (MOST 109-2112-M-007-037-MY3). SC would like to thank Aishrila Mazumder and Gourab Giri for fruitful discussions.
This research has made use of the data available in GMRT and VLA archives. We thank the staff of the GMRT who have made these observations possible. The GMRT is run by the National Centre for Radio Astrophysics of the Tata Institute of Fundamental Research. The National Radio Astronomy Observatory is a facility of the National Science Foundation operated under cooperative agreement by Associated Universities, Inc. We also thank the NRAO staff for making VLA observation possible. This research has used data obtained from the \textit{Chandra} Data Archive and the \textit{Chandra} Source Catalog and software provided by the \textit{Chandra} X-ray Center (CXC) in the application packages CIAO and Sherpa. This research made use of Astropy,\footnote{http://www.astropy.org} a community-developed core Python package for Astronomy \citep{Astropy_2013A&A...558A..33A,astropy_2018}, Matplotlib \citep{Matplotlib_Hunter:2007}, and APLpy, an open-source plotting package for Python \citep{APLpy_2012ascl.soft08017R}.

\section*{Data Availability}
The archival radio data used in our work are available in the GMRT
Online archive (\url{https://naps.ncra.tifr.res.in/goa/data/search}, project code 17\_019),the VLA Data Archive (\url{https://archive.nrao.edu/archive/advquery.jsp}, project codes AB0699, AO0149 and AM0469) and the Chandra data archive (\url{https://cda.harvard.edu/chaser/}, ObsId 15136)

\bibliography{sample631}{}
\bibliographystyle{aasjournal}

\appendix

\section{X-ray surface brightness profile over SE wedge region}
\begin{figure*}[h!]
\begin{center}
\begin{tabular}{lccr}
\includegraphics[width=0.33\textwidth]{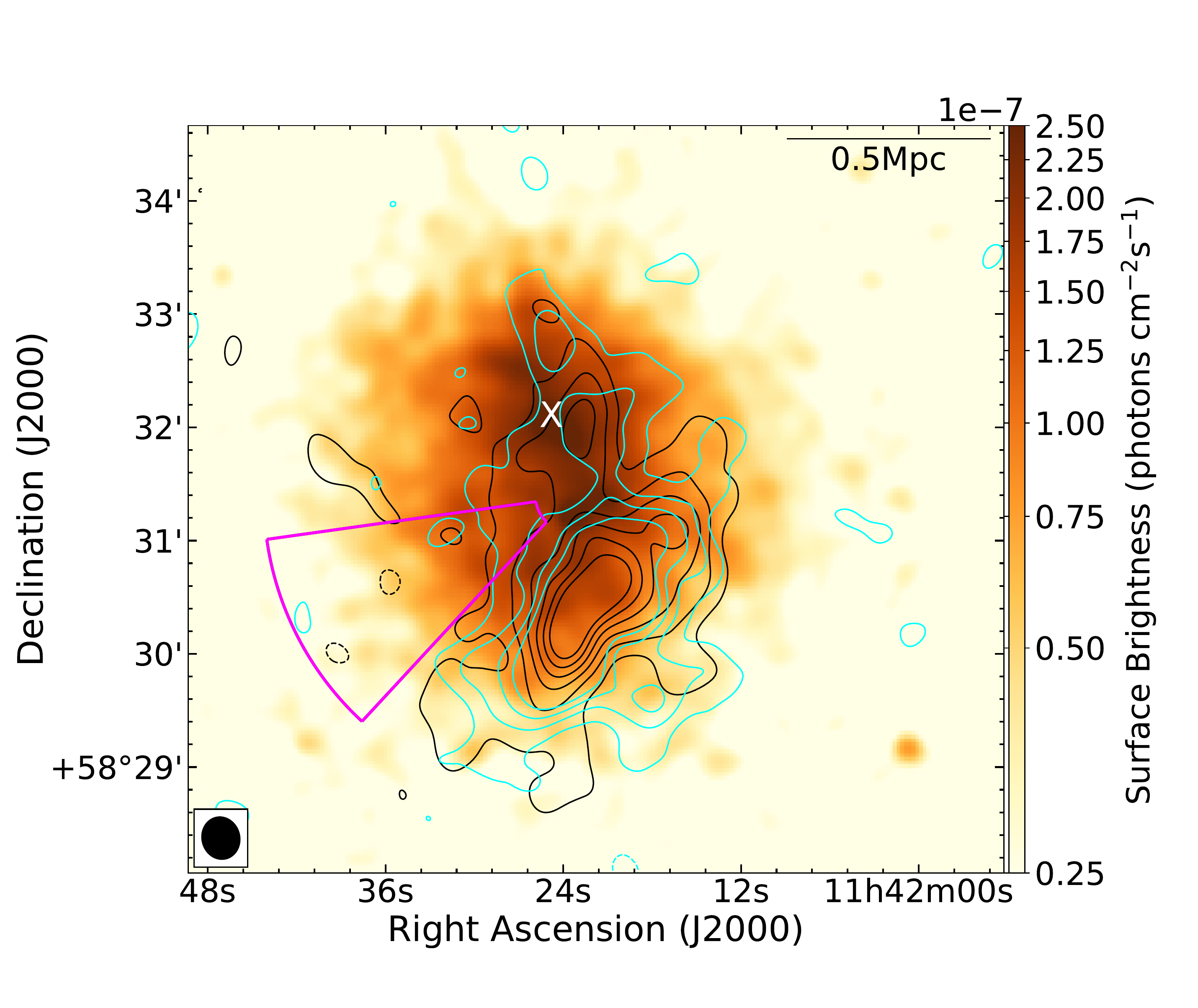} &
\includegraphics[width=0.30\textwidth]{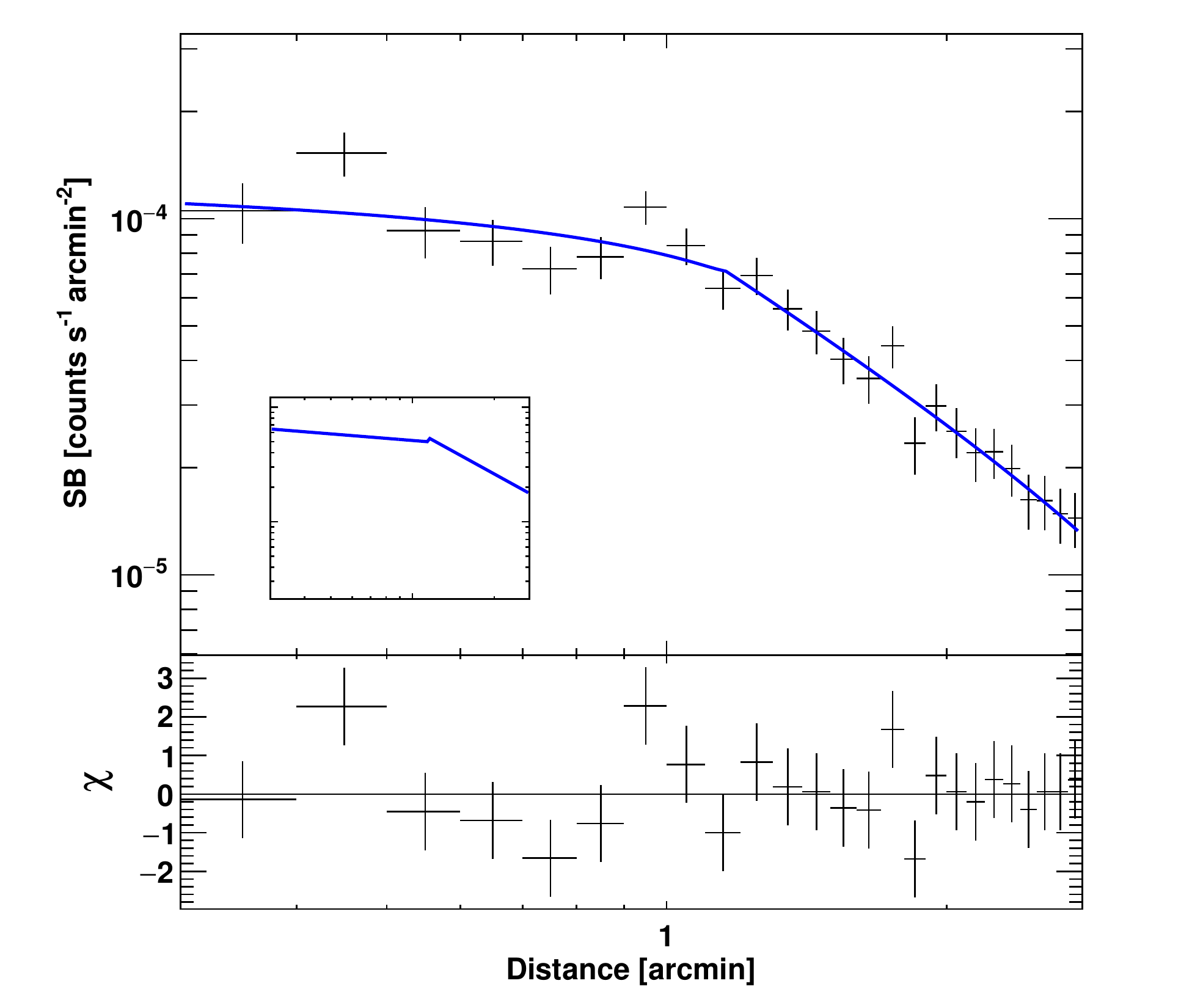} &
\includegraphics[width=0.30\textwidth]{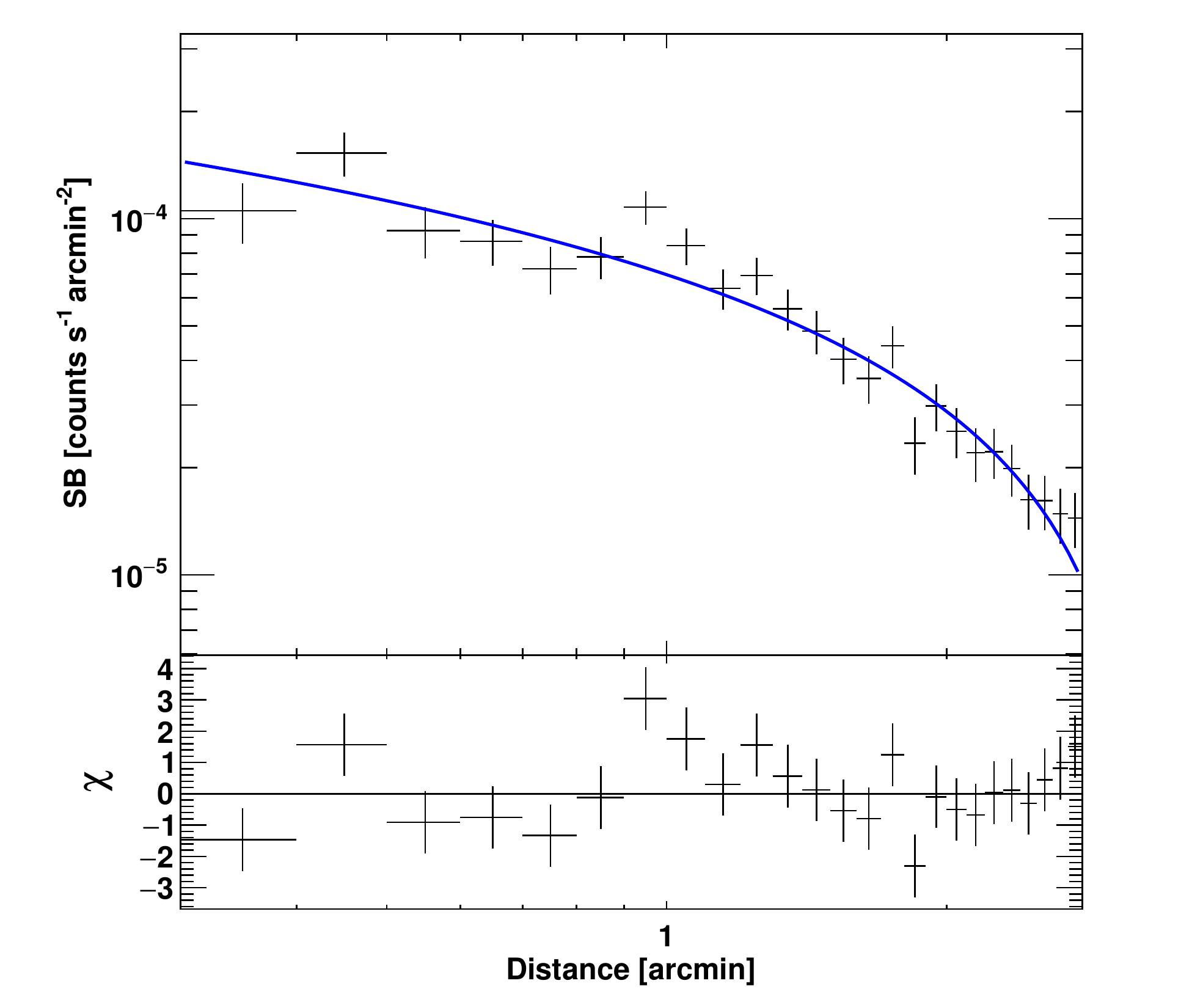} \\
\end{tabular}
\caption{Left: \textit{Chandra} X-ray surface brightness map of A1351 overlaid with 610  MHz GMRT low resolution image contours (black) and VLA 1.4  GHz low resolution image contours (cyan) with restoring beam $22.46\arcsec \times 19.88\arcsec$, P.A. $14.03\degree$.
The contours levels are placed at levels (-1,1,2,3,4,5,6) $\times\ 3\sigma_\mathrm{rms}$ where $\sigma_\mathrm{rms} = 260 \mu$Jy/beam for 610  MHz image and $\sigma_\mathrm{rms} = 50 \mu$Jy/beam for 1.4  GHz image. The wedge region (magenta) was used to produce a surface brightness profile. Middle: X-ray surface brightness profile obtained from the wedge region (magenta) shown in the left panel. This profile was fitted with a broken power law, but no discontinuity ($C = 0.92   \pm  0.15$) was detected. Right: X-ray surface brightnes profile fitted with Power law.}
\label{fig:se-plot}
\end{center}
\end{figure*}

To check the reliability of the surface brightness discontinuity detected in the southwest (SW) region of A1351, we created another similar profile along the southeast (SE) region (see Figure \ref{fig:se-plot}, left panel). We fitted the profile of this wedge region with broken power law (Figure \ref{fig:se-plot}, middle) but no density or temperature jump was observed over the wedge region. Figure \ref{fig:se-plot}, right panel shows a power law fit over the profile. The fitting parameters are mentioned in Table \ref{tab:Se-Table}. Unlike the SW profile, the SE profile shows no discontinuity. Therefore, we argue that the discontinuity in the SW profile (shown in Figure \ref{fig: xray shock}) might be caused by the merging shock.

\begin{table*}[h!]
	\centering
	\caption{Best fit parameters of broken power law model (using PROFFIT) are listed below.}
	\label{tab:Se-Table}
\resizebox{\textwidth}{!}{
	\begin{tabular}{lcccccr} 
		\hline
		\hline
		Model & $\alpha1$ & $\alpha2$ & $r_{sh}$(arcmin) & norm & Jump(C) & $\chi^2/D.o.f$ \\
		\hline
		Broken power law & 0.19$  \pm $0.18 & 1.30$   \pm $ 0.13 & 1.15$  \pm $0.01 & $5.11   \pm  0.82 e-05$ & 0.92 $  \pm $ 0.15 & 23.3/18 (1.29)  \\
		Power law & 6.27 $  \pm $ 3.29 e-02 & &  3.95 $  \pm $ 1.27 & 8.81 $  \pm $ 4.5 e-04 & & 32.37 / 20 (1.62) \\
		\hline
	\end{tabular}
}
\end{table*}

\end{document}